**Boron Arsenide Heterostructures: Lattice-Matched Heterointerfaces, and Strain Effects on Band Alignments and Mobility**

*Kyle Bushick, Sieun Chae, Zihao Deng, John Heron, and Emmanouil Kioupakis\**

Kyle Bushick, Sieun Chae, Zihao Deng, Prof. John Heron, Prof. Emmanouil
Department of Materials Science and Engineering, University of Michigan, Ann Arbor,
Michigan, 48109, USA
E-mail: kioup@umich.edu

BAs is III-V semiconductor with ultra-high thermal conductivity, but many of its electronic properties are unknown. This work applies predictive atomistic calculations to investigate the properties of BAs heterostructures, such as strain effects on band alignments and carrier mobility, considering BAs as both a thin film and a substrate for lattice-matched materials. The results show that strain decreases the band gap independent of sign or direction. In addition, biaxial tensile strain increases the in-plane electron and hole mobilities by more than 60% compared to the unstrained values due to a reduction of the electron effective mass and of hole interband scattering. Moreover, BAs is shown to be nearly lattice-matched with InGaN and $ZnSnN_2$, two important optoelectronic semiconductors with tunable band gaps by alloying and cation disorder, respectively. The results predict type-II band alignments and determine the absolute band offsets of these two materials with BAs. The combination of the ultra-high thermal conductivity and intrinsic p-type character of BAs, with its high electron and hole mobilities that can be further increased by tensile strain, as well as the lattice-match and the type-II band alignment with intrinsically n-type InGaN and $ZnSnN_2$ demonstrate the potential of BAs heterostructures for electronic and optoelectronic devices.

## 1. Introduction

Boron Arsenide (BAs) is an attractive electronic material due to its ultra-high thermal conductivity

(~1300 Wm$^{-1}$K$^{-1}$),[1–3] native p-type dopability,[4] and the availability of millimeter-size single crystals as substrates for thin-film growth.[3,5,6] Following the experimental validation[1–3] of the theoretical prediction[7,8] of its ultra-high thermal conductivity, research efforts have focused on its growth and fundamental characterization.[9] These results include studies of the thermal, optical, electronic, and structural properties of BAs in both bulk and 2D forms.[4,9–16] However, to fully evaluate its potential in electronic and optoelectronic applications, the properties of BAs heterostructures with other semiconductor materials, either as a thin film or as a substrate, must be investigated.

On one hand, BAs is a promising thin-film semiconductor. Two important degrees of freedom for thin-film engineering in device architectures are mechanical strain and band alignment. Strain arising from epitaxial mismatch strongly affects the electronic properties of materials, including the band gap,[17] band alignments,[18] effective masses,[19] and mobility.[20,21] For the epitaxial growth of BAs thin films on the (111) plane, candidate tetrahedrally bonded substrate materials are ZnO (3.249 Å[22]), GaN (3.189 Å[23]), or GaAs (3.997 Å[24]), which result in epitaxial strains ranging from –6% to +15% **(Figure 1(a))**. However, the effects of mechanical strain on the electronic properties of BAs are unknown.

On the other hand, BAs can also be utilized as a high-thermal-conductivity substrate for the epitaxial growth of other semiconductors **(Figure 1(b))**. Specifically, the lattice constant of the (111) plane of BAs (3.380 Å[1]) is well matched with InGaN alloys (3.189 – 3.533 Å[23,25,26]) and disordered ZnSnN$_2$ (~3.383 Å).[27] These materials are prominent optoelectronic materials due to their direct band gap in the visible range (and the resulting strong optical absorption and emission), which can be tuned by adjusting the alloy composition (InGaN) or the degree of cation disorder (ZnSnN$_2$),[28] but they both face challenges in device applications. On the one hand, the amount of In that can be incorporated into InGaN is limited in part due to the large mismatch with the underlying substrate, typically GaN or sapphire. On the other, ZnSnN$_2$ films are unintentionally heavily n-type in nature, and require a partner p-type semiconductor in devices such as LEDs or solar cells. Both materials could benefit from p-type substrates with minimal lattice misfit for high-quality thin-film growth. BAs can be a potential junction partner to these materials owing to its intrinsic p-type nature and lattice matching. However, band offsets for these hybrid structures and the assessment of their potential for device applications have not yet been determined.

In this work, we aim to understand the effect of biaxial strain on the band structure, mobility, and absolute band alignments of BAs with density functional theory (DFT) calculations and explore its possible applications in semiconductor heterostructures, either as a thin-film, where biaxial strain may enhance its electronic properties, or as a substrate, where it may form advantageous heterojunctions with other materials. We show that when epitaxially grown with 1% biaxial tensile strain, both electron and hole mobilities are increased by over 60%, showing promise for fast-switching and energy-efficient transistors. We also propose favorable nearly-lattice-matched heterojunctions of two important compounds, InGaN and ZnSnN$_2$, with BAs, showing how BAs can be used as a high-thermal-conductivity substrate for type-II band aligned heterojunctions for photovoltaic and photodetector applications.

## 2. Methodology

Our DFT calculations for the band structure and band alignment use the projector augmented wave (PAW) method[29,30] and the Heyd-Scuseria-Ernzerhof (HSE06)[31] hybrid functional as implemented in the Vienna *Ab initio* Simulation Package (VASP).[32–34] To construct biaxially strained structures, we first relax the cubic unit cell, fix the in-plane lattice constant ($a_\parallel$) under –4% to 4% strain, calculate the total energy as a function of the out-of-plane lattice constant ($a_\perp$) for each $a_\parallel$, identify the $a_\perp$ that minimizes the total energy **(Figure S1)**, and calculate the band structure of the resulting crystal structure including spin-orbit coupling effects. For band-alignment calculations, we build strained and unstrained BAs slabs consisting of 12 layers (48 atoms) along the nonpolar (110) direction and calculate the plane-averaged electrostatic potential without relaxing the ions. We then use the energy difference between the average electrostatic potential inside the slab and in the vacuum region to align the BAs bands to vacuum. For the $ZnSnN_2$ band alignment, we first align the order structure to the vacuum level using the same method as BAs except that the mixing parameter of HSE06 functional is chosen to be 31%, which is commonly used in previous DFT calculations on $ZnSnN_2$ to match the experimental band gap.[28,35] Then, the partially disordered $ZnSnN_2$ is aligned to the ordered structure directly using the band eigenvalues from the DFT calculations. The justification and details of this alignment procedure are provided in the Supporting Information.

To calculate the electron and hole mobilities in strained and unstrained BAs, we use the Quantum ESPRESSO[36] and Electron-phonon Wannier (EPW)[37,38] codes. We employ the local-density approximation (LDA) and the plane-wave norm-conserving pseudopotential method. We first relax the unit cell and subsequently apply a 1% biaxial tensile strain to the (001) plane and allow the out-of-plane (perpendicular) direction to relax to find the strained structure. We also use density functional perturbation theory[39] within Quantum ESPRESSO to calculate the phonon frequencies and displacements on a coarse 6×6×6 Brillouin-zone sampling (BZ-sampling) grid for the strained and unstrained geometries. We then use the EPW code to interpolate the electron-phonon coupling matrix elements to fine BZ-sampling meshes (up to 120×120×120) and determine the electron and hole mobilities at 300 K for the strained and unstrained structures for free-carrier concentrations of $10^{18}$ cm$^{-3}$. We also calculate the temperature dependence of mobility from 100 to 500 K with 50 K steps on fine BZ-sampling meshes of 72×72×72.

## 3. Results and Discussion

### 3.1. Strain Effects on the Band Structure and Absolute Band Positions

We first examine the effects of strain on the band structure along the Γ–X and Γ–Z directions, which include the conduction band minimum (CBM) and valence band maximum (VBM). These results, calculated with the HSE hybrid functional and spin-orbit effects, are shown in **Figure 2(a–b)** (**Figure S4** shows the LDA band structure without spin-orbit effects). For the unstrained structure, the X (in-plane) and Z (out-of-plane) directions are equivalent by symmetry, and the conduction band minima along the Γ–X and Γ–Z directions are degenerate (Figure 2(a)). However, biaxial strain within the (100) plane breaks the cubic symmetry and splits the degeneracy of the CBM valleys near X and Z; the lower CBM occurs along the Γ–Z (Γ–X) direction under tensile (compressive) strain (Figure 2(b)). The broken symmetry under strain also affects the VBM by splitting the topmost two degenerate valence bands (Figure 2(b)). These effects are important in explaining the mobility findings we discuss in the following paragraphs. Strain also shifts the absolute band positions as shown in **Figure 2(c)**. We find that the conduction band lowers and the valence band rises in energy with increasing strain (independent of its sign) within both the (100) and (111) planes. This decreases the band gap from 1.78 eV (unstrained) to 0.89 eV (for 4% tensile strain within the (100) plane) or 1.15 eV (for 4% compressive strain within the (100) plane). We find that strain within the (100) plane decreases the band gap more than within the (111) plane, probably due to the fact that the (100) plane is less close-packed than the (111) plane. Also, while tensile strain changes the band gap more than compressive strain for the (100) plane, the opposite trend is found for the (111) plane.

### 3.2. Strain Effects on Carrier Mobility

Next, we investigate the effect of tensile strain on the phonon-limited electron and hole mobility of BAs, i.e. the upper mobility limits in non-defective samples. We first evaluate the converged values by plotting the calculated mobility at 300 K versus the inverse of the number of the BZ-sampling points and extrapolating (**Figure S5**). For unstrained BAs we find electron and hole mobility values of 1341 and 1387 cm$^2$V$^{-1}$s$^{-1}$ at 300 K, respectively. We find excellent agreement between our calculated electron mobilities and the results of Liu et al.[13] (1400 cm$^2$V$^{-1}$s$^{-1}$). However, there are discrepancies for the hole mobility (2110 cm$^2$V$^{-1}$s$^{-1}$) since our mobility calculations do not include spin-orbit coupling, which results in a three-fold (instead of two-fold) degenerate VBM (Figure S4). This higher band degeneracy in the absence of spin-orbit effects results in more scattering channels and a proportionately lower mobility. In fact, our result of 1387 cm$^2$V$^{-1}$s$^{-1}$ is almost exactly 2/3 of the value from Liu et al.[13] (2110 cm$^2$V$^{-1}$s$^{-1}$), providing evidence for the strong impact of band degeneracy on the hole mobility. We note that our calculations for the electron mobility agree well with Liu et al.[13] since the CBM is not significantly affected by spin-orbit coupling.

We next examine the mobilities of the 1% tensile-strained structure. From these calculations we find that the in-plane and out-of-plane mobilities at 300 K for electrons are 2417 and 482 cm$^2$V$^{-1}$s$^{-1}$, respectively, while the corresponding values for holes are 3550 and 2956 cm$^2$V$^{-1}$s$^{-1}$. Therefore in BAs, tensile strain increases both the electron and hole mobilities of BAs, which is consistent with Si, which has a similar band structure.[20,21]

By examining the band structures in Figure 2(b), we gain useful insight into the different mechanisms contributing to the increased mobility. In the case of electrons, we observe that tensile strain results in the lifting of the 3-fold degeneracy in the conduction band **(Figure 3(a–b))**, with the out-of-plane minima occurring ~200 meV below the two in-plane minima for 1% strain (Figure 2(b)). This energy splitting has a strong effect on the effective mass of electrons. In an unstrained crystal, where all of the conduction valleys are degenerate (Figure 3(a)), the average effective mass of electrons is the directional average between the lighter transverse effective mass and the heavier longitudinal effective mass [$m^* = (2m_t^* + m_l^*)/3 = 0.52$].[11] On the other hand, in strained BAs, similar to strained silicon,[21] the majority of electrons occupy the lower-energy out-of-plane valleys (Figure 3(b)), so in-plane transport is governed only by the lighter transverse effective mass [$m^* = 0.24$],[11] while out-of-plane transport is governed only by the heavier longitudinal effective mass [$m^* = 1.09$].[11] This difference in effective mass therefore helps explain not only the elevated in-plane electron mobility, but also the suppressed out-of-plane electron mobility. Previous work on strained silicon also found that the changes in effective mass, as opposed to reduced intervalley scattering, are the dominant mechanism affecting the electron mobility.[21]

In the case of the hole mobility, the effective masses of the topmost three valence bands at Γ are all similar to one another, so the increase of the hole mobility by strain is explained by a different mechanism: the reduction in the number of available hole scattering channels **(Figure 3(c–f))**. Following from our previous discussion of the unstrained case, the inclusion of tensile strain leads to the topmost valence band becoming nondegenerate, with a separation of ~115 meV and ~145 meV with and without spin-orbit coupling, respectively. This reduction to only one hole-scattering-channel leads to an increase in the hole mobility by approximately a factor of two compared to the unstrained spin-orbit-coupling-corrected results (where there were two scattering channels). We note that our result for the strained hole mobility is not expected to be strongly affected by the inclusion of spin-orbit coupling, since the top valence band of strained BAs is non-degenerate – by a similar degree – whether spin-orbit coupling effects are included (Figure 3(f)) or not (Figure 3(e)). Unlike electrons, we find that strain also increases the out-of-plane hole mobility due to the same reduction in the number of hole scattering channels. We attribute the difference between the in-plane and out-of-plane hole-mobility values to the differences of the hole effective masses under strain, as the out-of-plane effective mass ($m^* = 0.25$) is slightly higher than the in-plane mass ($m^* = 0.22$). The qualitative conclusions that the strain increased hole mobility is due to the reduction of interband scattering caused by degeneracy lifting via broken symmetry, while the in-plane and out-of-plane hole-effective-mass differences are responsible for the slightly different values along the two orientations, are consistent with a similar investigation for strained Si.[21]

We also examine the temperature dependence of the electron and hole mobilities in the 100–500 K range **(Figure 4(a–b))**. For the unstrained case, we find that below 200 K, both the electron and hole mobilities follow a power law close to $T^{-1.5}$ **(Figure 4(c–d))**, indicating a primarily acoustic-deformation-potential carrier-scattering mechanism. From 200 to 500 K we find that the temperature dependence of the electron and hole mobilities of unstrained BAs follow a $T^{-2.14}$ and $T^{-2.41}$ power law, respectively (Figure 4(c–d)), which is similar to the values found by Liu et al.[13] and indicates that optical-deformation-potential scattering becomes the dominant carrier-scattering mechanism.[40] In the presence of tensile strain, acoustic deformation potential remains the dominant carrier-scattering mechanism below 200 K. But in the 200 to 500 K range the temperature dependence of the mobility changes compared to unstrained BAs – decreasing to $\sim T^{-1.8}$ for electrons and increasing to $\sim T^{-3.0}$ for holes (Figure 4(c–d)) – signifying a qualitative change in the dominant electron scattering mechanism by tensile strain. We attribute the different temperature dependence under tensile strain to the weakening (strengthening) of the optical-deformation-potential scattering of electrons (holes) due to the reduction of the valley (valence-band) degeneracy by tensile strain, respectively.

### 3.3. BAs as a Lattice-Matched Substrate

Next, we considered heterojunctions with other semiconductors where BAs is used as a substrate and their potential device applications. According to **Figure 5(a)**, we determine that BAs can form lattice-matched interfaces with InGaN and ZnSnN$_2$ and can solve several challenges that arise in adapting these materials for optoelectronic devices.

*3.3.1. InGaN*

First, InGaN alloys have emerged as prominent optoelectronic materials due to their efficient luminescence, high absorption coefficient, and the broad tunability of their direct band gap across the entire visible spectrum by alloying (from 0.69 eV for InN[41] to 3.4 eV for GaN[42]).[23,43,44] However, the difficulty in growing high-quality InGaN films with high In concentrations limits the emission wavelengths of devices to shorter than green and is one of the major challenges for fabricating efficient InGaN-based amber and red LEDs[44] for, e.g., full-color displays, or efficient solar cells.[45] This growth challenge arises mainly due to the lack of lattice-matched substrates. GaN-buffered sapphire (0001) is the most common substrate for InGaN growth, but the lattice mismatch between GaN and InGaN increases with increasing In composition. The calculated critical thickness for the formation of misfit dislocation decreases rapidly with increasing In composition and becomes < 1 nm for $In_{0.2}Ga_{0.8}N$/GaN heterostructures,[45] which is too thin for efficient light emission in LEDs and light absorption in solar cells.

Our calculations show that (111)-oriented BAs can be an alternative substrate or heterolayer for epitaxial InGaN growth. The hexagonal lattice constant of the BAs (111) plane (3.379 Å) is 5.62 % larger than GaN (3.189 Å) and 4.56 % smaller than InN (3.533 Å). Applying Vegard's law, we find that BAs can serve as an excellent substrate for $In_xGa_{1-x}N$ growth with misfit strain less than 1% for In compositions of ~ 0.47, with a band gap of 1.75 eV[46] that enables red light emission. Thus, by exploiting a BAs substrate or epilayer, the light-emission wavelength range of InGaN-based devices can be expanded to cover the full visible spectrum. Additionally, BAs and $In_xGa_{1-x}N$ (0.3 < x < 1) have type-II band alignment **(Figure 5(b))**.[47] Therefore, BAs substrates or epilayers can serve as a p-type junction for InGaN-based photovoltaic devices. For an In content of 50%, the band offsets between InGaN and BAs are calculated to be 0.42 eV for the CB and 0.30 eV for the VB, which allows for the effective separation and collection of carriers.

*3.3.2. ZnSnN$_2$*

Recently, the family of Zn-IV-N$_2$ compounds has been proposed as a next-generation class of solar-absorber materials composed of Earth-abundant elements that can replace the costly and rare In and Ga of conventional group-III nitrides.[27] Among their favorable characteristics, the tunability of their direct band gap (between 1.12 and 2.09 eV) at a fixed composition by adjusting the degree of cation disorder is of particular interest.[48] Thin-films of Zn-IV-N$_2$ compounds and alloys (e.g., ZnGe$_x$Sn$_{1-x}$N$_2$) have been successfully synthesized using molecular beam epitaxy[49] and sputtering,[50,51] and the tuning of their band gap through controlled cation disorder has been demonstrated by adjusting the growth temperature or relative ratio of source elements.[48] Despite their attractive properties, there are limited studies concerning device integration and characterization with these materials since intrinsically high electron concentrations from donor point defects (V$_N$, O$_N$, and Sn$_{Zn}$) hampers bipolar doping of this material.[35] Using alternative methods, a ZnSnN$_2$/p-Si heterojunction was fabricated,[52] while NiO, Cu$_2$O, and CuAlO$_2$ have also been proposed as other suitable p-type semiconductors.[53] Still, the lattice mismatch for all these materials ranges from 10% to 15%, which is too large to achieve high-crystal-quality epitaxial interfaces.

We propose n-ZnSnN$_2$/p-BAs as an optimal hybrid structure for efficient photovoltaic applications. First, the cation disordered pseudo-wurtzite structure of ZnSnN$_2$ has very close lattice match along the (111) plane of BAs with misfit strain less than 1%. Also, while ZnSnN$_2$ is intrinsically n-type, BAs is p-type due to B-related acceptor type point defects and its doping level can also be controlled extrinsically by Si, Ge, and Be acceptors.[4] Band alignment of the heterojunction in the intrinsic case is shown in Figure 5(b) with band offset energies between BAs and ordered ZnSnN$_2$ of 0.516 eV for the VB and 0.503 eV for the CB. Thus, effective separation of electrons and holes into different regions of the device is predicted. Furthermore, we note that the ideal pair of band gaps for a two-gap multi-junction photovoltaic device are 1.1 eV and 1.8 eV. While both BAs and ZnSnN$_2$ materials have intrinsic gaps close to 1.8 eV, a 1.1 eV gap in ZnSnN$_2$ may be achieved by full cation disorder, or a 1.1 eV gap in BAs may be achieved by epitaxial strain up to 4%, as shown in Figure 2(c).

## 4. Conclusion

In conclusion, we examine the potential of BAs for electronic and optoelectronic device applications by predicting important electronic properties of BAs for incorporation into devices as both a substrate material and an epitaxially grown thin film. As a thin film, we find that tensile epitaxial strain increases the in-plane electron and hole mobilities to 2417 and 3550 cm$^2$V$^{-1}$s$^{-1}$, respectively. These values represent an 80% and 68% increase over the unstrained case, a significant improvement that could lead to faster switching and more energy efficient transistors. We find that at temperatures below 200 K acoustic deformation potential dominates carrier scattering, while above 200 K optical deformation potential becomes the dominant scattering mechanism. For bulk BAs as a substrate, we find In-rich InGaN and ZnSnN$_2$ are candidate materials that can be epitaxially grown on BAs and form p-n junctions for lasers, LEDs, photodetectors, or photovoltaic cells. These junctions can cover the full visible spectrum, while simultaneously benefiting from the high thermal conductivity of BAs substrates for efficient heat removal.

**Acknowledgements**
K. B. and S. C. contributed equally to this work. We thank Dr. Samuel Poncé and Kelsey Mengle for their assistance with the EPW package and mobility calculations. This work was supported by the Designing Materials to Revolutionize and Engineer our Future (DMREF) Program under Award No. 1534221, funded by the National Science Foundation. This research used resources of the National Energy Research Scientific Computing Center, a DOE Office of Science User Facility supported by the Office of Science of the U.S. Department of Energy under Contract No. DE-AC02-05CH11231.

References

[1] J. S. Kang, M. Li, H. Wu, H. Nguyen, Y. Hu, *Science.* **2018**, *361*, 575.
[2] S. Li, Q. Zheng, Y. Lv, X. Liu, X. Wang, P. Y. Huang, D. G. Cahill, B. Lv, *Science.* **2018**, *361*, 579.
[3] F. Tian, B. Song, X. Chen, N. K. Ravichandran, Y. Lv, K. Chen, S. Sullivan, J. Kim, Y. Zhou, T.-H. Liu, M. Goni, Z. Ding, J. Sun, G. Amila, G. Udalamatta, H. Sun, H. Ziyaee, S. Huyan, L. Deng, J. Zhou, A. J. Schmidt, S. Chen, C. Chu, P. Y. Huang, D. Broido, L. Shi, G. Chen, Z. Ren, *Science.* **2018**, *361*, 582.
[4] S. Chae, K. Mengle, J. T. Heron, E. Kioupakis, *Appl. Phys. Lett.* **2019**, *113*, 212101.
[5] T. L. Chu, A. E. Hyslop, *J. Appl. Phys.* **1972**, *43*, 276.
[6] J. Xing, X. Chen, Y. Zhou, J. C. Culbertson, J. A. Freitas, E. R. Glaser, J. Zhou, L. Shi, N. Ni, *Appl. Phys. Lett.* **2018**, *112*, 261901.
[7] L. Lindsay, D. A. Broido, T. L. Reinecke, *Phys. Rev. Lett.* **2013**, *111*, 025901.
[8] T. Feng, L. Lindsay, X. Ruan, *Phys. Rev. B* **2017**, *96*, 161201.
[9] F. Tian, Z. Ren, *Angew. Chemie Int. Ed.* **2018**, *58*, 5824.
[10] N. K. Ravichandran, D. Broido, *Nat. Commun.* **2019**, *10*, 827.
[11] K. Bushick, K. Mengle, N. Sanders, E. Kioupakis, *Appl. Phys. Lett.* **2019**, *114*, 022101.
[12] J. L. Lyons, J. B. Varley, E. R. Glaser, J. A. Freitas, J. C. Culbertson, F. Tian, G. A. Gamage, H. Sun, H. Ziyaee, Z. Ren, *Appl. Phys. Lett.* **2018**, *113*, 251902.
[13] T.-H. Liu, B. Song, L. Meroueh, Z. Ding, Q. Song, J. Zhou, M. Li, G. Chen, *Phys. Rev. B* **2018**, *98*, 081203.
[14] H. Şahin, S. Cahangirov, M. Topsakal, E. Bekaroglu, E. Akturk, R. T. Senger, S. Ciraci, *Phys. Rev. B* **2009**, *80*, 155453.
[15] K. Manoharan, V. Subramanian, *ACS Omega* **2018**, *3*, 9533.
[16] S. Ullah, P. A. Denis, M. G. Menezes, F. Sato, *Appl. Surf. Sci.* **2019**, *493*, 308.
[17] Q. Yan, P. Rinke, A. Janotti, M. Scheffler, C. G. Van De Walle, *Phys. Rev. B* **2014**, *90*, 125118.
[18] C. M. Jones, E. Kioupakis, *J. Appl. Phys.* **2017**, *122*, 045703.
[19] C. E. Dreyer, A. Janotti, C. G. Van De Walle, *Appl. Phys. Lett.* **2013**, *102*, 142105.
[20] S. Poncé, D. Jena, F. Giustino, *arXiv:1908.02072v1*.
[21] D. Yu, Y. Zhang, F. Liu, *Phys. Rev. B* **2008**, *78*, 245204.
[22] H. Karzel, W. Potzel, M. Köfferlein, W. Schiessl, M. Steiner, U. Hiller, G. M. Kalvius, D. W. Mitchell, T. P. Das, P. Blaha, K. Schwarz, M. P. Pasternak, *Phys. Rev. B* **1996**, *53*, 11425.
[23] J. Wu, *J. Appl. Phys.* **2009**, *106*, 011101.
[24] M. E. Straumanis, C. D. Kim, *Acta Crystallogr.* **1965**, *19*, 256.
[25] M. Hori, K. Kano, T. Yamaguchi, Y. Saito, T. Araki, Y. Nanishi, N. Teraguchi, A. Suzuki, *Phys. Status Solidi Basic Res.* **2002**, *234*, 750.
[26] M. Kurouchi, T. Araki, H. Naoi, T. Yamaguchi, A. Suzuki, Y. Nanishi, *Phys. Status Solidi Basic Res.* **2004**, *241*, 2843.
[27] N. Senabulya, N. Feldberg, R. A. Makin, Y. Yang, G. Shi, M. J. Christina, E. Kioupakis, J. Mathis, R. Clarke, S. M. Durbin, *AIP Adv.* **2016**, *6*, 075019.
[28] R. A. Makin, K. York, S. M. Durbin, N. Senabulya, J. Mathis, R. Clarke, N. Feldberg, P. Miska, C. M. Jones, Z. Deng, L. Williams, E. Kioupakis, R. J. Reeves, *Phys. Rev. Lett.* **2019**, *122*, 256403.
[29] P. E. Blöchl, *Phys. Rev. B* **1994**, *50*, 17953.

[30]  D. Joubert, *Phys. Rev. B* **1999**, *59*, 1758.
[31]  J. Heyd, G. E. Scuseria, M. Ernzerhof, *J. Chem. Phys.* **2003**, *118*, 8207.
[32]  G. Kresse, J. Hafner, *Phys. Rev. B* **1993**, *47*, 558.
[33]  G. Kresse, J. Furthmüller, *Comput. Mater. Sci.* **1996**, *6*, 15.
[34]  G. Kresse, J. Furthmüller, *Phys. Rev. B - Condens. Matter Mater. Phys.* **1996**, *54*, 11169.
[35]  S. Chen, P. Narang, H. A. Atwater, L. Wang, *Adv. Mater.* **2013**, *26*, 1.
[36]  P. Giannozzi, S. Baroni, N. Bonini, M. Calandra, R. Car, C. Cavazzoni, D. Ceresoli, G. L. Chiarotti, M. Cococcioni, I. Dabo, A. Dal Corso, S. De Gironcoli, S. Fabris, G. Fratesi, R. Gebauer, U. Gerstmann, C. Gougoussis, A. Kokalj, M. Lazzeri, L. Martin-Samos, N. Marzari, F. Mauri, R. Mazzarello, S. Paolini, A. Pasquarello, L. Paulatto, C. Sbraccia, S. Scandolo, G. Sclauzero, A. P. Seitsonen, A. Smogunov, P. Umari, R. M. Wentzcovitch, *J. Phys. Condens. Matter* **2009**, *21*, 395502.
[37]  F. Giustino, M. L. Cohen, S. G. Louie, *Phys. Rev. B* **2007**, *76*, 165108.
[38]  S. Poncé, E. R. Margine, C. Verdi, F. Giustino, *Comput. Phys. Commun.* **2016**, *209*, 116.
[39]  S. Baroni, S. de Gironcoli, A. Dal Corso, P. Giannozzi, *Rev. Mod. Phys.* **2001**, *73*, 515.
[40]  D. K. Ferry, *Phys. Rev. B* **1976**, *14*, 1605.
[41]  J. Wu, W. Walukiewicz, W. Shan, K. M. Yu, J. W. Ager, S. X. Li, E. E. Haller, H. Lu, W. J. Schaff, *J. Appl. Phys.* **2003**, *94*, 4457.
[42]  I. Vurgaftman, J. R. Meyer, *J. Appl. Phys.* **2003**, *94*, 3675.
[43]  J. Wu, W. Walukiewicz, K. M. Yu, W. Shan, J. W. Ager, E. E. Haller, H. Lu, W. J. Schaff, W. K. Metzger, S. Kurtz, *J. Appl. Phys.* **2003**, *94*, 6477.
[44]  S. A. Kukushkin, A. V Osipov, V. N. Bessolov, B. K. Medvedev, V. K. Nevolin, K. A. Tcarik, *Rev. Adv. Mater. Sci.* **2008**, *17*, 1.
[45]  A. G. Bhuiyan, K. Sugita, A. Hashimoto, A. Yamamoto, *IEEE J. Photovolt.* **2012**, *2*, 276.
[46]  P. G. Moses, M. Miao, Q. Yan, C. G. Van de Walle, *J. Chem. Phys.* **2011**, *134*, 084703.
[47]  P. G. Moses, C. G. Van de Walle, *Appl. Phys. Lett.* **2010**, *96*, 021908.
[48]  T. D. Veal, N. Feldberg, N. F. Quackenbush, W. M. Linhart, D. O. Scanlon, L. F. J. Piper, S. M. Durbin, *Adv. energy Mater.* **2015**, *5*, 1501462.
[49]  P. C. Quayle, K. He, J. Shan, K. Kash, *MRS Commun.* **2013**, *3*, 135.
[50]  L. Lahourcade, N. C. Coronel, K. T. Delaney, S. K. Shukla, N. A. Spaldin, H. A. Atwater, *Adv. Mater.* **2013**, *25*, 2562.
[51]  F. Deng, H. Cao, L. Liang, J. Li, J. Gao, H. Zhang, R. Qin, C. Liu, *Opt. Lett.* **2015**, *40*, 1282.
[52]  R. Qin, H. Cao, L. Liang, Y. Xie, F. Zhuge, H. Zhang, J. Gao, K. Javaid, C. Liu, W. Sun, *Appl. Phys. Lett.* **2016**, *108*, 142104.
[53]  A. N. Fioretti, Development of Zinc Tin Nitride for Application as an Earth Abundant Photovoltaic Absorber, Colorado School of Mines, 2017.

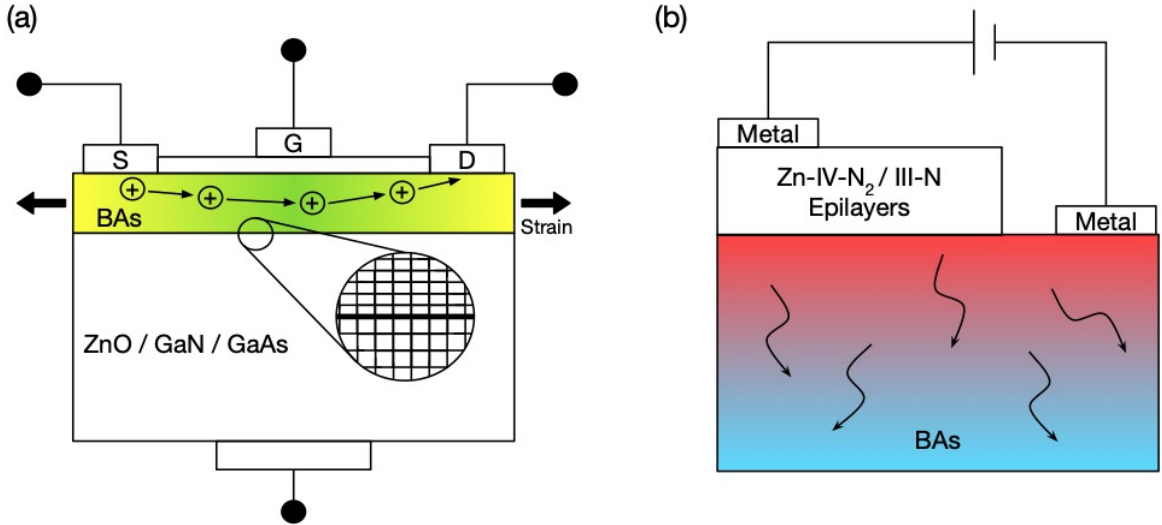

**Figure 1.** Potential configurations of BAs in semiconductor devices. (a) Epitaxially grown BAs thin film as part of a transistor. Epitaxially straining BAs (inset) increases both the electron and the hole mobility. (b) Schematic of an optoelectronic device utilizing the junction between a thermally conducting BAs substrate for efficient heat extraction in conjunction with nearly-lattice-matched direct-band-gap semiconductor films for efficient light emission and absorption.

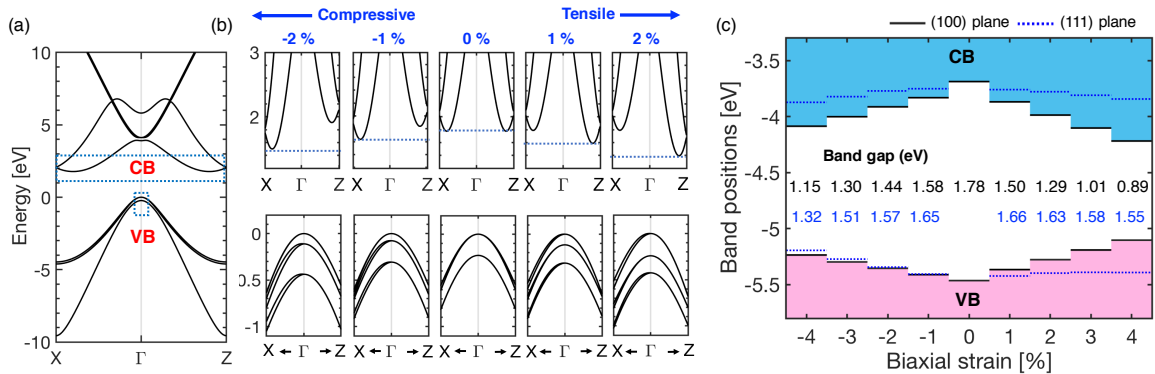

**Figure 2.** (a) Calculated band structure of unstrained BAs along X to Γ and Γ to Z using HSE hybrid functional considering spin-orbit effects. The X and Z directions are equivalent due to the cubic symmetry. (b) The effect of biaxial strain within the (100) plane on the band structure of BAs. The upper panel shows the splitting of the degeneracy of the CBM valleys near X and Z under strain due to the broken symmetry. Tensile strain places electrons at the Z valley while compressive moves them to the X valleys. The bottom row shows the splitting of the degenerate topmost valence bands under strain. (c) Absolute band alignments of BAs relative to the vacuum level as a function of biaxial strain within the (100) (black lines) and (111) (blue lines) planes. The band gap is always reduced independent of the strain direction or sign.

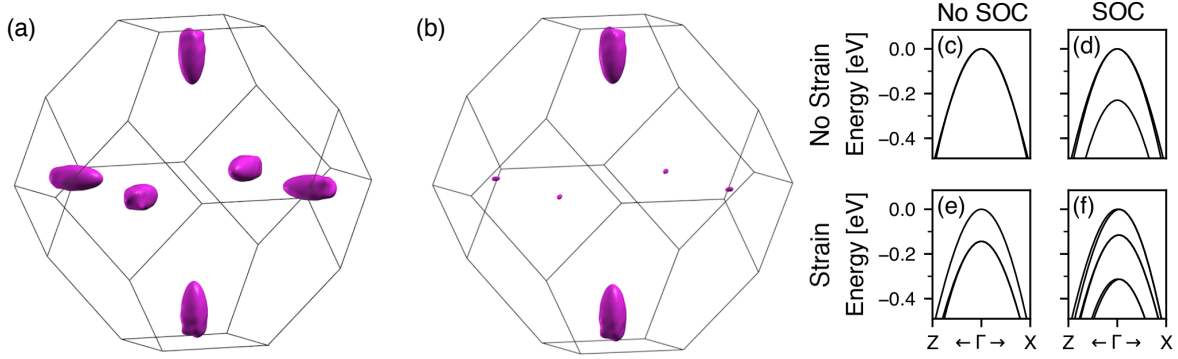

**Figure 3.** Isosurfaces of the lowest conduction band at an energy 200 meV above the CBM in the first BZ for (a) unstrained and (b) 1% tensile-strained BAs. In the unstrained case all six valleys are symmetrically equivalent, while in the strained case carriers primarily populate the lower-energy out-of-plane valleys. In the case of holes, the Figure shows the top valence bands for (c) no strain or spin-orbit coupling (3-fold degenerate VBM), (d) no strain but with spin-orbit coupling (2-fold degenerate VBM), (e) 1% tensile strain but no spin-orbit coupling (non-degenerate VBM), and (f) 1% tensile strain and spin-orbit coupling (non-degenerate VBM). The calculated hole mobilities are primarily governed by the number of channels (bands) that contribute to hole scattering.

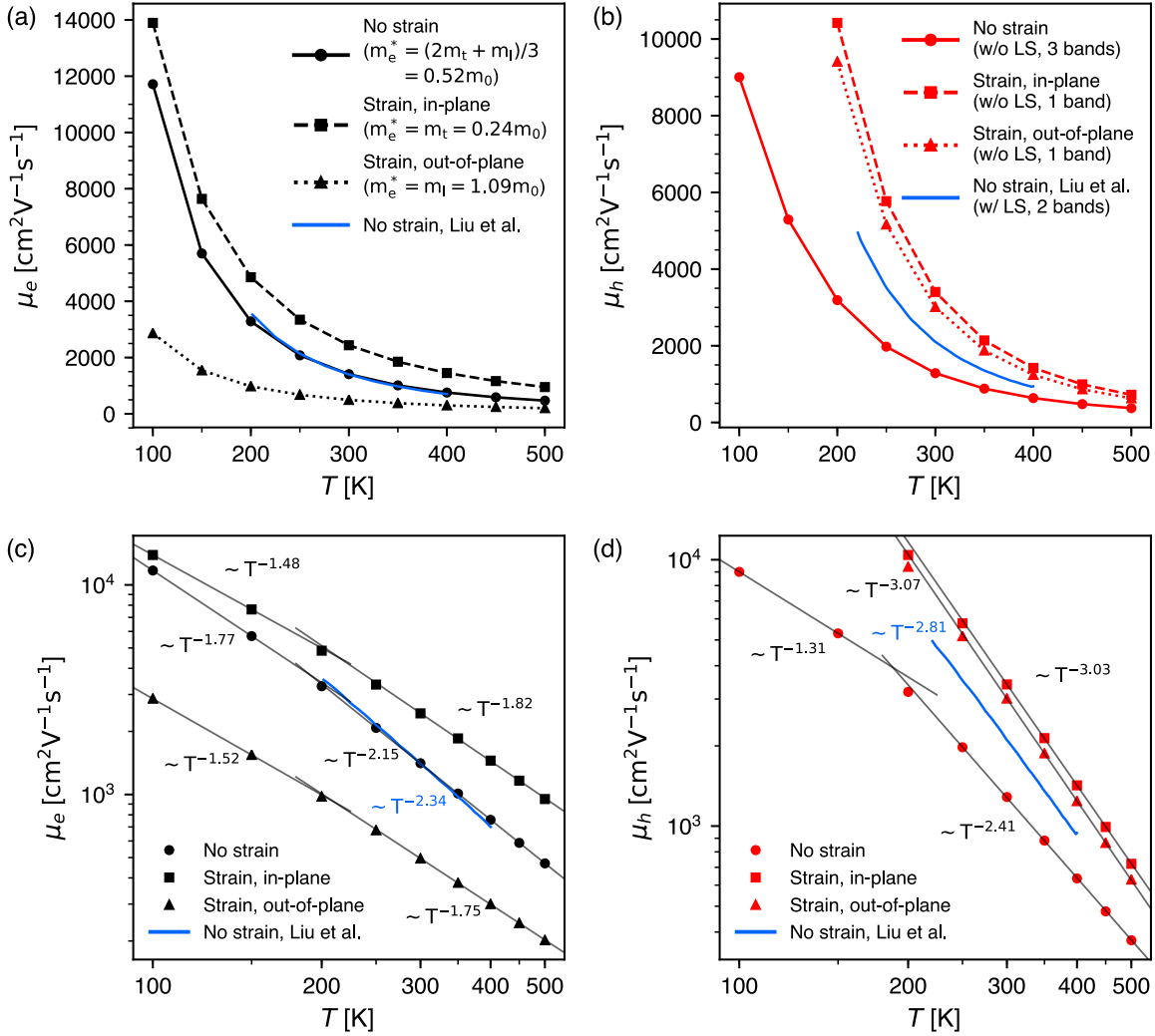

**Figure 4.** Temperature dependence of (a) electron and (b) hole carrier mobilities of BAs under tensile strain. The solid lines with circle markers show the results for unstrained BAs from our calculations (black/red) and those of Liu et al. (blue).[13] The mobilities under strain are shown for both the in-plane (dashed/squares) and out-of-plane (dotted/triangles) directions. Tensile strain increases the in-plane and reduces the out-of-plane electron mobility (due to changes to the effective mass), while it also increases both the in-plane and out-of-plane hole mobility (by breaking the topmost valence band degeneracy and reducing interband scattering). Temperature dependence of electron (c) and hole (d) mobilities shown on a log-log scale. There is a clear change in the power law governing the temperature dependence around 200 K in all cases. In the case of holes (and to a lesser extent electrons), strain increases the magnitude of the temperature dependence, indicating a change in the dominant scattering mechanism. Note that data from this work do not include spin-orbit coupling.

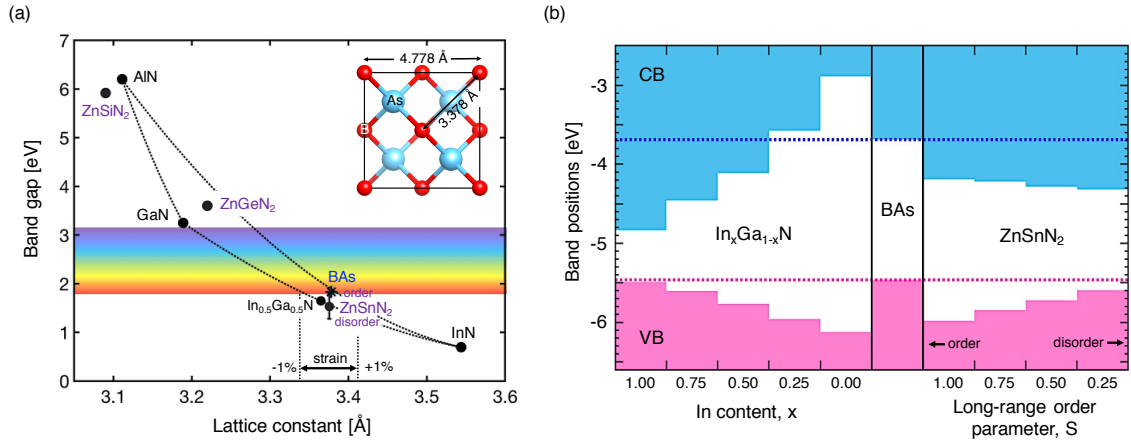

**Figure 5.** (a) Band gap versus in-plane lattice constant for BAs, wurtzite group-III nitrides, and orthorhombic Zn-IV-nitrides. $In_xGa_{1-x}N$ with In concentration (x) from 0.46 to 0.65 and $ZnSnN_2$ are nearly lattice-matched to BAs with misfit strain less than 1%. (b) Band alignments of unstrained $In_xGa_{1-x}N$, BAs, and $ZnSnN_2$ with respect to the vacuum level. The band positions of $In_xGa_{1-x}N$ are plotted as a function of In content, x, while those of $ZnSnN_2$ are plotted as a function of the long-range order parameter, S. BAs has type-II band alignments with n-type $In_{0.5}Ga_{0.5}N$ and $ZnSnN_2$, showing BAs can be a potential p-junction partner with these materials for efficient solar cell absorber. The band alignment data for InGaN is adapted from ref.[47]

# Supporting Information

Boron Arsenide Heterostructures: Lattice-Matched Heterointerfaces, and Strain Effects on Band Alignments and Mobility

*Kyle Bushick, Sieun Chae, Zihao Deng, John Heron, and Emmanouil Kioupakis**

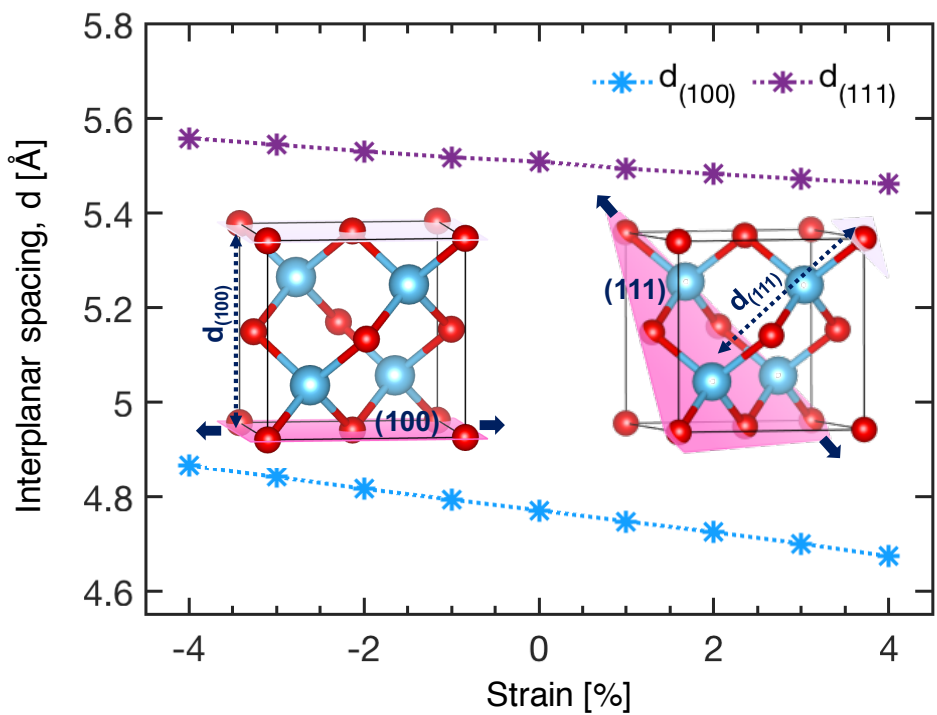

**Figure S1.** The out-of-plane interplanar distance of BAs as a function biaxial strain when the strain is applied within (100) and (111) planes.

**Band alignment of order and disorder ZnSnN$_2$**

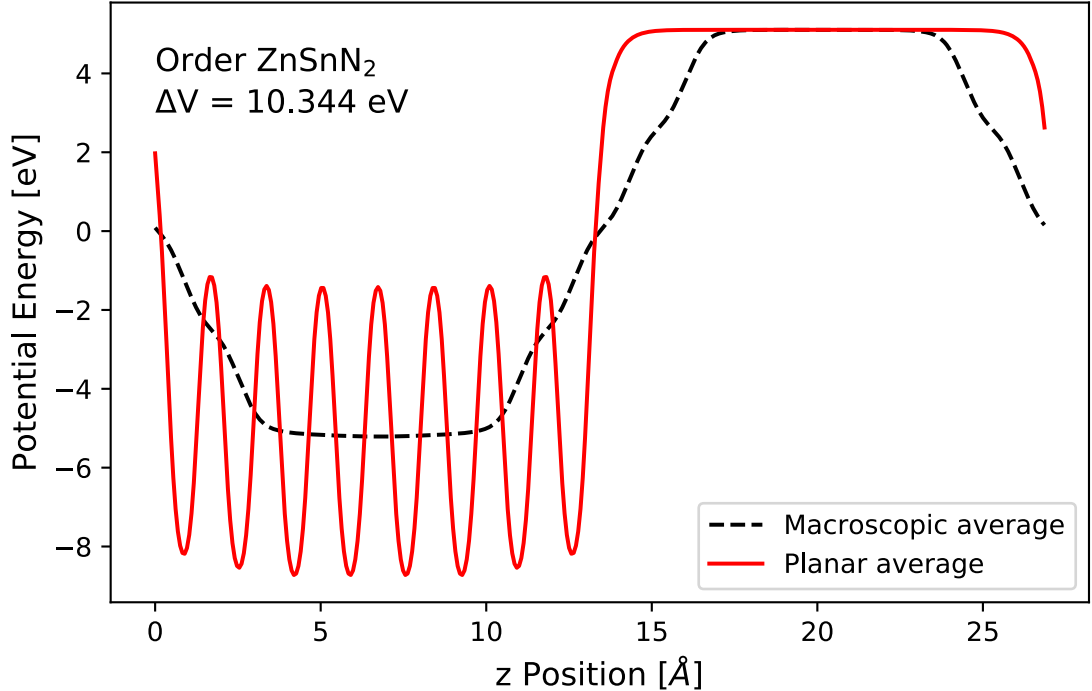

**Figure S2.** Planar-averaged electrostatic potential of ordered ZnSnN$_2$ along the wurtzite nonpolar ($11\bar{2}0$) direction (z direction). The macroscopic electrostatic potential is averaged over one unit cell along the z direction. The difference in electrostatic potential between the bulk and the vacuum region is 10.334 eV.

To perform the ZnSnN$_2$ band alignment calculations, we first relax the ordered structure and calculate its band gap by using the HSE06 hybrid functional[1] with a mixing parameter of 31%. Then an 8-layer slab (32 atom) is built along the wurtzite nonpolar ($11\bar{2}0$) direction, and its average electrostatic potential is calculated with HSE06 without relaxing the ions (See Figure S4). The absolute band positions are determined from the energy difference of electrostatic potential between the bulk and the vacuum region.

To model disordered ZnSnN$_2$, we built 2×2×2 supercells (128 atoms) with random placement of Zn and Sn on the cation sublattice based on the long-range order parameter S. We then select the most thermodynamically favorable structures by choosing the supercells that exclude the Zn4/Sn4 motifs, and maximize the fraction of Zn2Sn2 motifs. This follows the calculations by Lany et al.[2], who found that Zn4/Sn4 motifs are too energetically unfavorable to form. Subsequently, these supercells are fully relaxed, and their band gaps are calculated using the HSE06 functional. The valence band maximum (VBM) is determined from the average of the top 3 occupied valence states, while the conduction band minimum (CBM) is determined from the lowest unoccupied state.

The relative band alignment between the disordered structures and the ordered one is determined directly from the VBM, and CBM eigenvalues from the DFT calculations. This is because the energy differences of the electrostatic potential between the bulk and the vacuum region are almost the same with respect to the cation disorder in $ZnSnN_2$ (See Figure S5). These electrostatic potentials are calculated using the slabs constructed along the wurtzite $(10\bar{1}0)$ directions, which ensures the vacuum regions are nonpolar. Due to the computational cost, these calculations are performed with PBE-GGA functional[3].

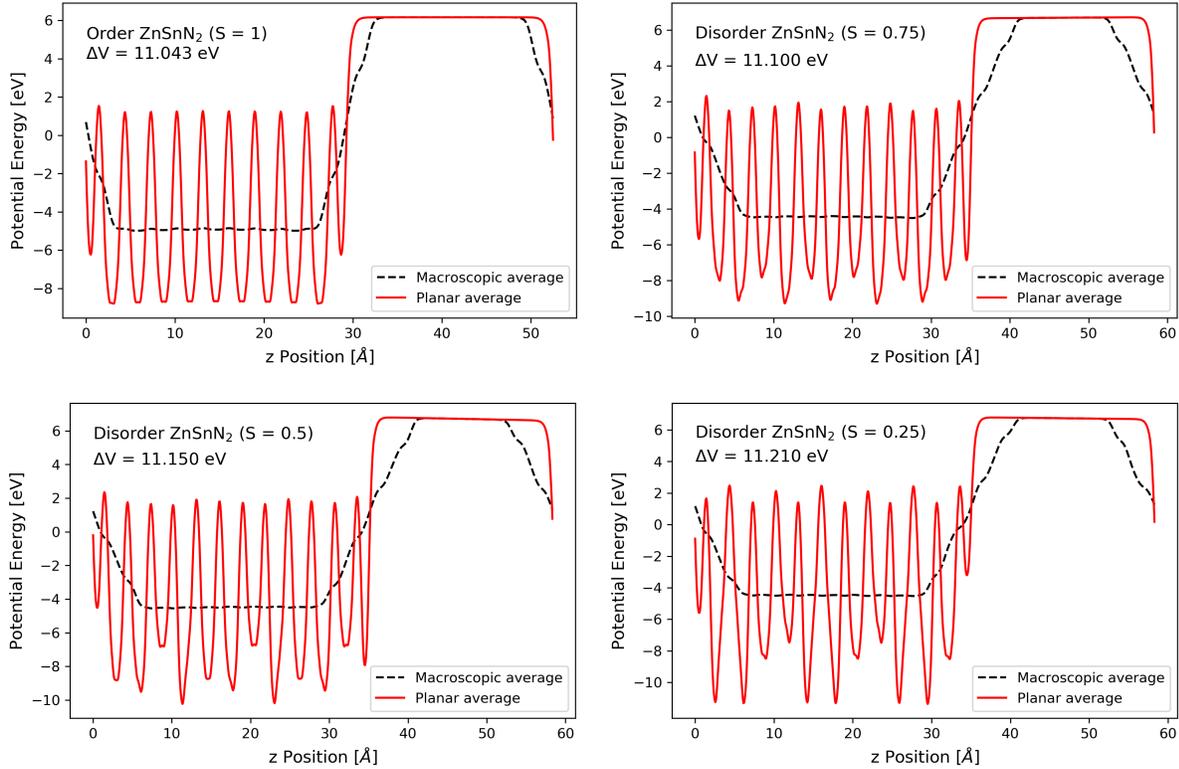

**Figure S3.** Planar average electrostatic potential of order and disorder $ZnSnN_2$ along the wurtzite nonpolar $(10\bar{1}0)$ direction (z direction). The macroscopic electrostatic potential is average over one supercell along the z direction. The difference in electrostatic potential between the bulk and the vacuum region varies within 150 meV for different degree of disorder, which indicates that disorder does not change the position of the bulk average electrostatic potential to the vacuum level.

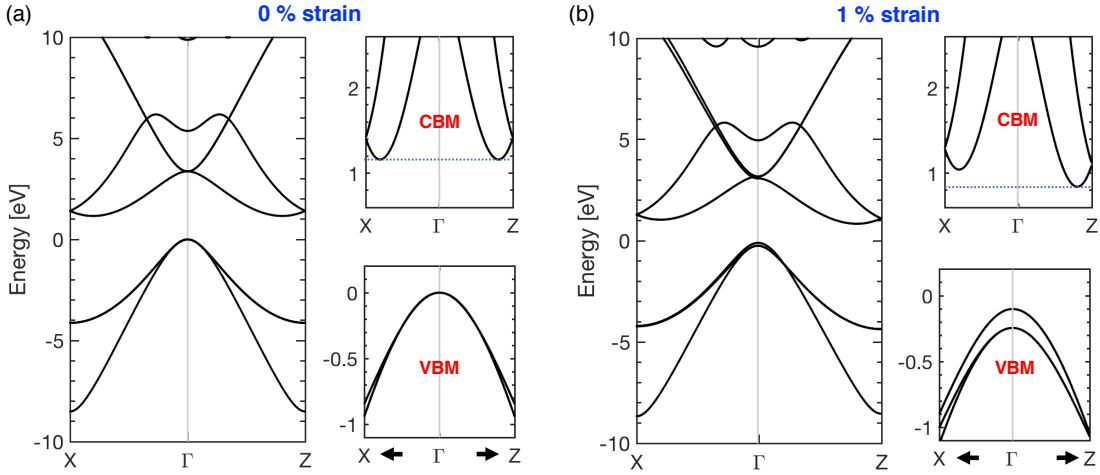

**Figure S4.** Calculated band structure of (a) unstrained and (b) 1 % tensile strained BAs along X to $\Gamma$ and $\Gamma$ to Z using an LDA functional without spin-orbit effects.

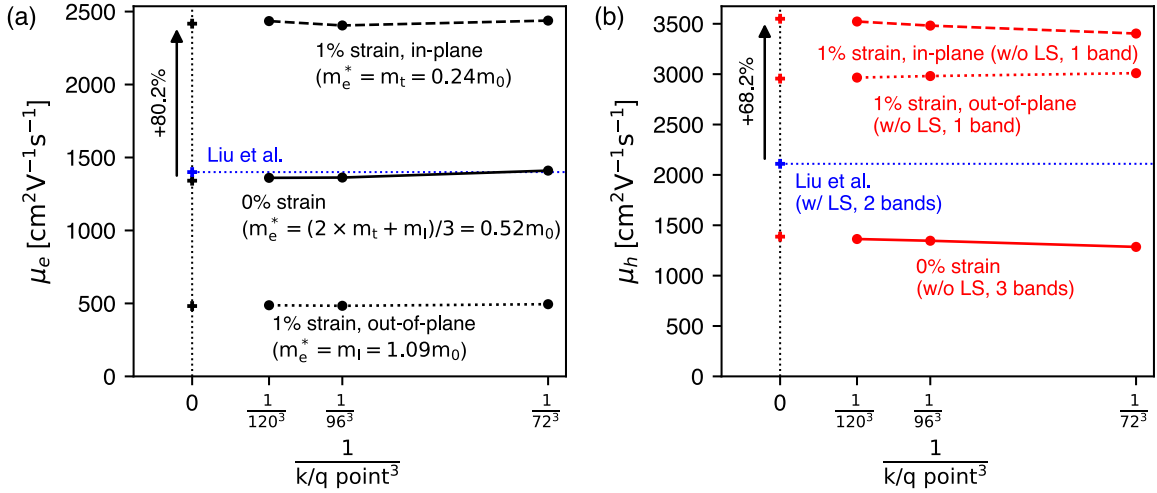

**Figure S5.** Convergence of (a) electron and (b) hole mobilities calculated without spin-orbit coupling at 300 K as a function of increasing electron and phonon Brillouin-zone sampling-grid density. We extrapolate to the fully converged value and compare to previously reported values from Liu et al. (blue dotted line).[4] We find that the in-plane electron mobility (dashed) increases by 80% under 1% tensile strain due to the lighter transverse effective mass, whereas the out-of-plane mobility (dotted) is reduced by 64%. Strain also increases the hole mobility by 68% in-plane (dashed) and 40% out-of-plane (dotted) due to the reduction of the number of bands that holes scatter to.

References

[1]  J. Heyd, G. E. Scuseria, M. Ernzerhof, *J. Chem. Phys.* **2003**, *118*, 8207.
[2]  S. Lany, A. N. Fioretti, P. P. Zawadzki, L. T. Schelhas, E. S. Toberer, A. Zakutayev, A. C. Tamboli, *Phys. Rev. Mater.* **2017**, *1*, 035401.
[3]  J. P. Perdew, K. Burke, M. Ernzerhof, *Phys. Rev. Lett.* **1996**, *77*, 3865.
[4]  T.-H. Liu, B. Song, L. Meroueh, Z. Ding, Q. Song, J. Zhou, M. Li, G. Chen, *Phys. Rev. B* **2018**, *98*, 081203.